\documentclass[twocolumn,preprintnumbers,elsart]{revtex4}
\usepackage{makeidx}
\usepackage{amssymb}
\usepackage{amsmath}
\usepackage{mathrsfs}
\usepackage{graphicx}
\usepackage{dcolumn}
\usepackage{bm}
\usepackage[center]{subfigure}
\usepackage{color}

\begin{document}

\title{Femtosecond soliton diode on heterojunction Bragg-grating structure}
\author{Zhigui Deng$^{1}$, Haolin Lin$^{1}$, Hongji Li$^{2}$, Shenhe Fu$^{3}$, Yikun Liu$^{2}$, Ying Xiang$^{4}$, Yongyao Li}
\email{yongyaoli@gmail.com}
\affiliation{$^{1}$Department of Applied Physics, College of Electronic Engineering, South China Agricultural University, Guangzhou 510642, China\\
$^{2}$State Key Laboratory of Optoelectronic Materials and Technologies,\\
Sun Yat-sen University, Guangzhou 510275, China\\
$^{3}$Department of Optoelectronic Engineering, Jinan University, Guangzhou 510632, China\\
$^{4}$ School of Information Engineering, Guangdong University of Technology, Guangdong University of Technology, Guangzhou 510006, China}

\begin{abstract}
We numerically propose a scheme for realizing an all-optical femtosecond soliton diode based on a tailored heterojunction Bragg grating, which is designed by two spatially asymmetric chirped cholesteric liquid crystals. Our simulations demonstrate that with the consideration of optical nonlinearity, not only the femtosecond diode effect with nonreciprocal transmission ratio up to 120 can be achieved, but also the optical pulse evolving into soliton which maintains its shape during propagation through the sample is observed. Further, the influence of pulse width and the carrier wavelength to the femtosecond diode effect is also discussed in detail. Our demonstrations might suggest a new direction for experimentally realizing the femtosecond soliton diode based on the cholesteric liquid crystals.
\end{abstract}
\maketitle

Diode, which allows the nonreciprocal transportation of electric current, is one of the most important electronic devices in the integrated circuit. In optics, the nonreciprocal effect also plays a significant role in the modern information processing as well as in optically integrated devices. However, optical light beam is usually reciprocal and therefore how to transmit optical signal only in one direction and block that in the opposite direction remains a challenging issue. \\
\indent In the past decades, many schemes were proposed to solve this problem. The most significant proposal was based on the optical isolator realized by the Faraday effect in magneto-optical crystals \cite{Carson1964}. Since the magneto-optical isolator requires a strong external pump, it hinders the applications in a highly integrated photonic systems. Other schemes were also reported with different principles or materials such as using opto-acoustic effect \cite{Kang2011}, second-harmonic generation \cite{Carlos1996}, absorbing multilayer system \cite{Gevorgyan2002}, asymmetric nonlinear absorption \cite{Philip2007}, ring-resonator structures \cite{Fan2009,Fan2012,Fan2015}, Fano effect \cite{Wding2012,Xuyi2014}, micro-optomechanical devices with flexible Fabry-Perot \cite{Lipson2009}, left-hand materials \cite{Feise2005}, photonic crystal structures \cite{Wang2013,Zhang2014} and parity-time system \cite{Chang2014}. We note that although high nonreciprocal transmission ratio could be achieved using these schemes, the shape of the light field after propagating through the devices is usually neglected. Due to diffraction/dispersion as well as the interaction with materials, the light beam might cannot maintain its profile, which would limit the application in optical processing. Recently, linearly-coupled waveguide system in which the unidirectional propagation are expected in the nonlinear regime \cite{Khawaja2015} was reported. Based on the fiber Bragg grating (FBG), a sandwich structure was proposed to realize optical nonreciprocal effect \cite{Hongji2015}. We should mention that with the consideration of optical nonlinearity, these optical waves \cite{Khawaja2015,Hongji2015} could maintain their profiles during propagation through the nonlinear devices. Indeed, the light wave propagating though these nonlinear systems is in a form of solitons which can preserve its shape during propagation \cite{BG1,BG2,BG3,BG4,BGT,Atai2001,Fu1,Fu2}. \\
\indent In this Letter, we extend the concept of optical diode from picosecond \cite{Hongji2015} to femtosecond (fs) domain for the first time, which is of particular interest in the ultrafast data processing. However, the realization of fs diode becomes more difficult than that realized in the picosecond domain. The reason lies in the fact that the fs pulse contains a much boarder spectrum that the bandgap of majority of the photonic crystals, e.g., FBG, cannot cover. This is because the depth of the refractive index modulation in FBG is very small (10$^{-2}$$\sim$10$^{-3}$ \cite{Caohui2012}), giving rise to a narrow photonic bandgap. Recently, it has been shown that cholesteric liquid crystals (CLCs), which is a natural BG structure with refractive index modulation depth near 10$^{-1}$ \cite{SongLY}, can be used to stretch and compress the fs pulses \cite{Yikun2016}. Importantly, a large nonlinear coefficient in the order of 10$^{-12}$ cm$^2/$W with femtosecond response was reported in CLCs \cite{SongLY, Yikun2016,Hwang2004,Hwang2005-2}. These findings open a new avenue to realize optical diode in femtosecond domain. \\
\indent Here, we propose a tailored heterojunction Bragg grating, which is consisted of two spatially asymmetric chirped cholesteric liquid crystals for realizing soliton diode in femtosecond domain. We should mention that the proposed scheme is different from that reported in the sandwich FBG structure \cite{Hongji2015} and the photonic bandgap liquid-crystal heterojunctions \cite{Hwang2005}. We note that the chirped CLCs, which is equivalent to the chirped Bragg-gratings (CBGs) in FBG \cite{CBG}, can be practically achieved by the experiment \cite{Ilchishin2011}. Our numerical simulations based on the standard nonlinear coupled mode theory (CMT) suggest that this setting can support the unidirectional transportation of the fs solitons.
\begin{figure}[tbp]
\centering{\label{fig1a} \includegraphics[width=6cm]{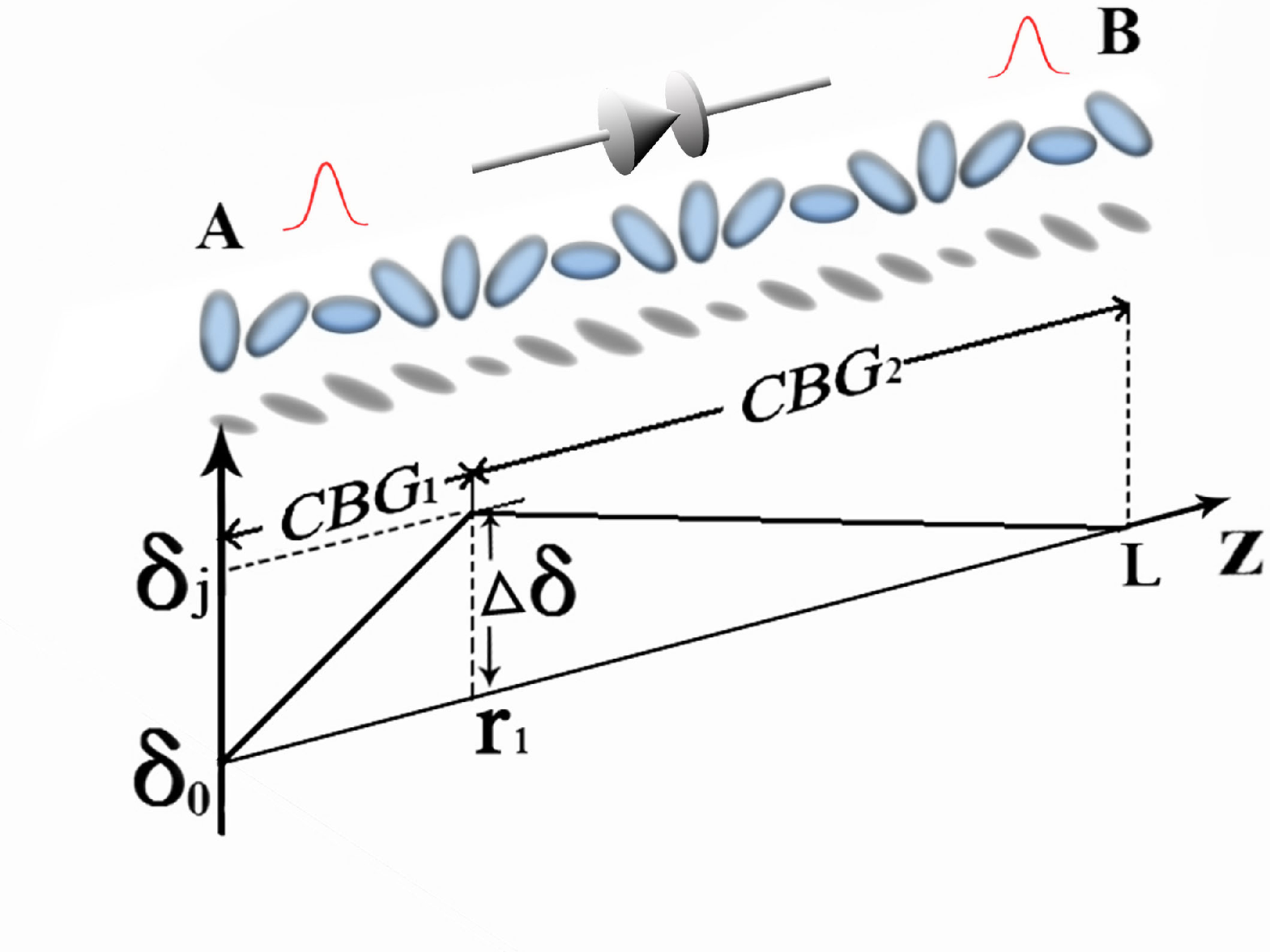}}
\caption{The sketch of the tailored structure, which is consisted of two spatially asymmetric chirped Bragg grating (CBG) based on the CLCs. The junction between the two CBGs is located at $z=r_{1}$. The blue ellipsoid denotes the orientations of the CLC molecules. Note that CBG$_{1,2}$ represent the left and right chirped CLC segments, respectively.}
\label{sketch}
\end{figure} \\
\indent CMT has been proven suitable for describing the dynamic of fs pulses through the CLCs \cite{SongLY,Yikun2016}. In this Letter, we adopt the CMT, and therefore the pulses propagating through the setting are described by the following standard nonlinear coupled-mode equations (NCMEs):
\begin{eqnarray}
&&\pm i{\partial \over\partial z}E_{f,b} +{i\over v_g}
{\partial  \over\partial t}E_{f,b}+
\delta(z)E_{f,b}+\kappa E_{b,f}\nonumber\\
&&+\gamma(|E_{f,b}|^{2}+2|E_{b,f}|^{2})E_{f,b}=0, \label{basicEq}
\end{eqnarray}
where $t$ is time, $E_f$ and $E_b$ are amplitudes of the forward-traveling and backward-traveling waves of the light field.  It was reported that the photonic bandgap structure is exhibited provided that the circularly polarized light is launched in CLCs \cite{Yikun2016}. $v_g=c/n_0$ is the linear group velocity in CLCs, with $c$ being the light speed in the vacuum and $n_{0}=1.635$. $\gamma=n_{2}\omega/c$ is the nonlinear parameter, with $\omega$ being the carrier frequency and $n_{2}=-1.5\times10^{-11}$ cm$^{2}$/W the local self-defocusing Kerr coefficient (as the nonlinear response is an ultrafast process, the thermal nonlinearity of CLCs could be neglected \cite{SongLY,Yikun2016}). The coupling strength between forward and backward waves is $\kappa=\pi\Delta n/\lambda_{B}$, where $\Delta n=0.19$ is the depth of the refractive-index modulation and $\lambda_{B}=2n_{0}\Lambda_{0}$ is the Bragg wavelength at the input edge of the sample, with $\Lambda_{0}=238.53$ nm being the BG period which is exactly equal to 1/2 pitch of the CLCs at the input and output edges. The above parameters are selected from the experiment in Ref. \cite{SongLY}. $\delta(z)$, which represents the local wavenumber detuning of the chirp CLCs, is given by
\begin{eqnarray}
\delta(z)=\begin{cases}
\delta_{0}-2\pi C_{1}z/\Lambda_{0} & z\in[0,r_{1}],\\
\delta_{0}-2\pi C_{2}(L-z)/\Lambda_{0} & z\in(r_{1},L],
\end{cases} \label{deltaz}
\end{eqnarray}
where $L$ is the total length of the sample, and $r_{1}$ is the length of the chirped CLC segment on the left side, hence $L-r_{1}$ is the length of the right side of the sample. $\delta_{0}\equiv2\pi n_{0}/\lambda - \pi/\Lambda_{0}$ is the detuning at the left and right edges of the sample, where $\lambda$ is the incident wavelength of the light field. $C_{1}=\Delta \delta\Lambda_{0}/2\pi r_{1}$ and $C_{2}=\Delta \delta\Lambda_{0}/2\pi (L-r_{1})$ are the chirped coefficients of the left and right chirped CLC segments, respectively. Here, $\Delta \delta=\delta_{0}-\delta_{j}$ is the detuning difference at the junction of the two chirped CLC segments with $\delta_{j}$ being the detuning at the junction position. The sketch of the designed setting is displayed in Fig. \ref{sketch} Note that the location of the junction is at $z=r_{1}$. \\
\indent In simulations, we assume the total length of the sample is $L=500$ $\mu$m, which is in the same scale with that reported in \cite{Yikun2016}. It is worth mentioning that to achieve such chirped CLCs with hundreds of micrometer is difficult, as the focal conic texture and the absorption of dye-agents limit the thickness of the CLCs. In principle, these difficulties might be solved by the following processes: firstly, increasing the thickness of CLC film to hundreds of millimeter is needed. To avoid the occurrence of highly scattering focal conic texture, an additional electric field or magnetic field should be used, which could stabilize the planar texture, hence obtaining a high transparency; secondly, to produce CLCs with flexible pitch and length, the absorption of dye should be minimized by decreasing the doping concentration as low as possible while enlarging the exposure time to compensate this dilution.  Besides, as the CLCs feature a self-defocusing Kerr nonlinearity, we select the incident wavelength that is located at the red side of the photonic bandgap. It indicates that the detuning, i.e. $\delta(z)$, is negative. Therefore, according to previous studies \cite{Fu1, Hongji2015}, to see the soliton diode effect, the detuning difference, $\Delta \delta$, should be set positively.
\begin{figure}[tbp]
\centering{\label{fig2a} \includegraphics[width=8.5cm]{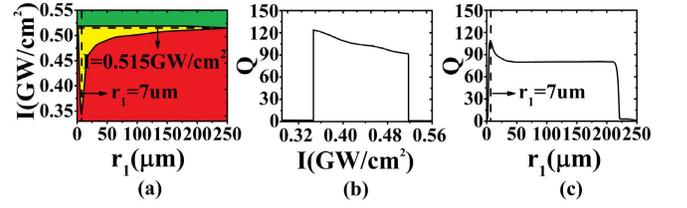}}
\caption{(a) Three transmission types of femtosecond pulses, completely cut-off (red color), diode-effect (yellow color) and breakdown-effect (green color) in the $I-r_{1}$ plane. The two black dashed lines show the optimized cross-section for the diode effect respectively. $r_{1}=7$ $\mu$m denotes the vertical dashed line; while $I=0.515$ GW/cm$^{2}$ denotes the horizontal dashed line. (b,c) The Q-factor of the femtosecond diode for the cases of the cross-sections shown in the black dash lines in panel (a). In all the panels, $\Delta\delta\equiv\delta_{j}-\delta_{0}=1055.3$ cm$^{-1}$. The temporal width of the pulse is 100 fs.}
\label{Qfactor}
\end{figure}\\
\indent Numerical studies of Eq. (\ref{basicEq}) are performed with a femtosecond Gaussian pulse, having a temporal width of 100 fs and a tunable peak intensity, namely $I$. According to the above discussion, the carrier wavelength of the pulse is selected as $\lambda=829$ nm, which is located at the red side of the photonic bandgap. Simulations are carried out by means of the 4-step Runge-Kutta method. Obviously, three types of transmission of femtosecond pulse through the setting, i.e., completely cut-off, diode-effect and breakdown-effect are found, displayed in the $I-r_{1}$ plane [see Fig. \ref{Qfactor}(a)]. In the cut-off region in which case the intensity-induced nonlinearity is insufficient, the pulses are bounced back from both sides; see the red region. As a result, the femtosecond soliton cannot penetrate the sample. While in the breakdown region; see the green region, due to the strong induced nonlinearity, the pulse can propagate through the setting from both sides. Particularly, with suitable incident pulse intensity, nonreciprocal transmission of the pulses are found; see the yellow region of Fig. 2(a). In this case, the solitons can propagate from left to right hand side but it bounces back when it is injected from the right edge of the setting. This phenomenon has been explained in Ref. \cite{Hongji2015}: the propagation dynamics of solitons can be approximately considered as a quasi-particle propagating in the setting, which cannot exhibit unidirectional transmission in an effective potential. However, a Gaussian optical pulse that is not an exact solution of the soliton and cannot use the quasi-particle approximation during the time when the pulse is developing into a soliton. The developing length (from a Gaussian pulse to a BG soliton) depends on the chirped coefficient of CLCs segment as well as the initial peak intensity of pulse. Therefore, an asymmetric chirped CLC pair produces asymmetric developing lengths of each side, thus the diode region is created when appropriate chirped coefficients of the setting and peak intensity of the pulse are selected. We should point out that due to the periodic modulation, the initial Gaussian pulse splits into reflected and transmitted pulses at the incident edge of the sample, but the transmitted pulse could eventually evolve into soliton since the propagating length is much larger than the dispersion length ($\sim 10 \mu$m). The region of diode effect closes up when $r_{1}=250$ $\mu$m, above which the system becomes reciprocal. From Fig. 2(a), we also see that there exists optimized incident pulse intensity $I=0.515$ GW/cm$^{2}$ and length of the left side chirped CLC $r_{1}=7$ $\mu$m, for which it gives the largest region that produces the diode effect; see the black dashed lines, respectively. \\
\indent To measure the quality of femtosecond soliton diode, a quality factor of the diode is defined by
\begin{eqnarray}
Q=\eta_{\mathrm{AB}}/\eta_{\mathrm{BA}}, \label{Q}
\end{eqnarray}
where $\eta_{\mathrm{AB}}$ and $\eta_{\mathrm{BA}}$ are the transmissivity of the light field from the left side to the right side (i.e. A$\rightarrow$B) and vice versa (i.e. A$\leftarrow$B), which are defined by
\begin{eqnarray}
\eta_{\{\mathrm{AB,BA}\}}={\int[|E_{f}(z_{\{L,0\}},t)|^{2}+|E_{b}(z_{\{L,0\}},t)|^{2}]dt\over\int[|E_{f}(z_{\{0,L\}},t)|^{2}+|E_{b}(z_{\{0,L\}},t)|^{2}]dt}. \label{eta}
\end{eqnarray}
The measured Q-factors for the cases of two optimized cross-section shown in dashed line in Fig. 2(a) are displayed in Figs. \ref{Qfactor}(b,c), respectively. Figures 2(b) and 2(c) show that the Q-factor with value larger than 80 can be achieved, and the maximum is up to 120, indicating a high nonreciprocal effect. A typical example of the diode effect is displayed in Fig. \ref{Exp}, showing the propagation dynamics of femtosecond solitons. It is clearly seen that the shape of the pulse is maintained during evolution through the setting. Note that the result shown in Fig. 3 is based on the parameters that are selected from the vertical dashed line in Fig. \ref{Qfactor}(a). In this case, the transmissivity from A$\rightarrow$B is $\eta_{\mathrm{AB}}=$ 34.2\%, while the transmissivity from A$\leftarrow$B is $\eta_{\mathrm{BA}}=$ 0.3\%. Note that again owing to the photonic bandgap, most of the pulse energy is reflected from the edge of the sample. Further, the pulse group velocity of the soliton is also measured to be 0.14$c$ by the formula $V_{g}=L/t=500 \mu\mathrm{m}/11.7 \mathrm{ps}$. It suggests that the femtosecond soliton can be also used as an optical buffer.
\begin{figure}[tbp]
\centering{\label{fig3a} \includegraphics[width=7.5cm]{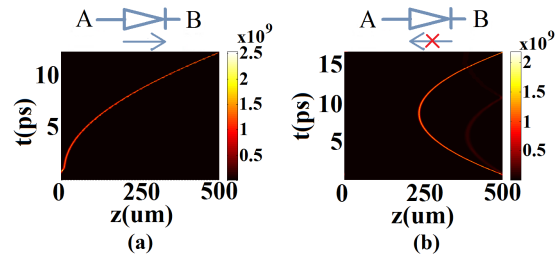}}
\caption{Typical example of the femtosecond diode effect. The pulse width is selected as 100 fs, and the pulses are injected from left (a) and right (b) hand sides, respectively. The related parameters are selected from that shown in the dashed lines in Fig. 2(a). The incident peak intensity of the pulse is $I=0.35$ GW/cm$^{2}$. (a) The femtosecond soliton can propagate from A to B; (b) the femtosecond soliton bounces back when it is injected from B. }
\label{Exp}
\end{figure}
\begin{figure}[tbp]
\centering{\label{fig4a} \includegraphics[width=8.5cm]{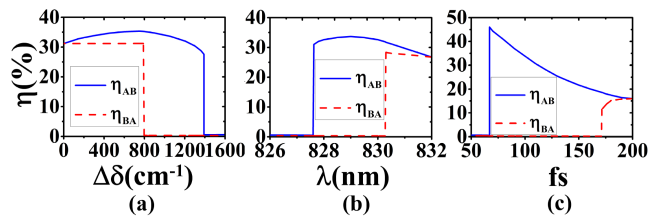}}
\caption{ The transmissivity $\eta_{\mathrm{AB}}$ (blue solid curves) and $\eta_{\mathrm{BA}}$ (red dashed curves) versus detuning difference (a), the incident carrier wavelength of pulses (b), and the pulse width (c). In panel (a) the wavelength and the width of the pulse are set as 829 nm and 100 fs, respectively. In panel (b), the detuning difference of the setting is $\Delta\delta=1055.3$ cm$^{-1}$ and the width of the pulse is 100 fs. In panel (c) the detuning difference of the setting is $\Delta\delta=1055.3$ cm$^{-1}$ and the wavelength of the pulse is 829 nm. In all cases, $r_{1}=7$ $\mu$m, and the peak intensity of the pulses is fixed to $I=0.43$ GW/cm$^{2}$.}
\label{transmissivity}
\end{figure} \\
\indent Next, we also investigate the influence of the detuning difference, $\Delta \delta$ (see Fig. \ref{deltaz}), to the femtosecond soliton diode. We note that $\Delta \delta$ plays an important role in building the nonreciprocal transmission of the system. Figure \ref{transmissivity}(a) displays the transmissivity in two incident cases versus $\Delta\delta$ for a fixed value of $r_{1}$ and pulse intensity. Remind that the incident wavelength of the light pulse remains unchanged. As can be seen, when $\Delta\delta=0$, the relation $\eta_{\mathrm{AB}}\equiv\eta_{\mathrm{BA}}$ can be obtained as the whole setting returns to a uniform BG structure. However, increasing $\Delta\delta$, the diode effect occurs. The diode effect disappears when $\Delta\delta$ is too large, e.g., when $\Delta\delta>$1390 cm$^{-1}$, we have $\eta_{\mathrm{AB}}\approx\eta_{\mathrm{BA}}\approx0$, indicating that the solitons are all reflected from both sides. This is because increasing the value of $\Delta\delta$ gives rise to shifting the carrier wavelength to the center of the bandgap at the junction location. As a result, the reflection of pulse energy becomes larger. \\
\indent Finally, we still study the influence of different wavelength and pulse width to the diode effect. The related results are shown in Figs. 4(b) and 4(c), respectively. Figure 4(b) shows that when 827.6 nm$<\lambda<$830.3 nm, significant diode effect still can be observed. $\lambda=827.6$ is the low-cut-off wavelength of the femtosecond diode, below which the pulse cannot penetrate into the setting owing to the limited incident pulse intensity; while $\lambda=830.3$ is the up-cut-off wavelength, above which the soliton is completely transmitted and the diode effect disappears as the wavelength of the soliton is far away from the center of the bandgap. Figure 4(c) represents the dependence of $\eta$ on the pulse width. Our numerical result demonstrates that the femtosecond diode effect is valid with the pulse width ranging from 67 fs to 171 fs. \\
\indent The objective of this work was to propose a scheme for all-optical diode in femtosecond domain. Our scheme was based on two chirped cholesteric liquid crystals (CLCs) linked by a heterojunction. The spatially asymmetric chirped CLCs provides opportunity for the nonreciprocal propagation of femtosecond solitons. The propagation dynamics of the femtosecond soliton through the setting can be described by the standard nonlinear coupled mode equations. Our numerical simulations demonstrated that diode effect can be supported by this setting. Particularly, optimized windows for diode effect were found. In the optimized windows, high nonreciprocal transmission ratio up to 120 was obtained based on this setting. Also, we still discussed the influence of pulse temporal width and carrier wavelength to the diode effect. Our numerical simulations still suggested that this setting can be applied to an ultrafast optical buffer, or power switch, etc. \\
\indent Zhigui Deng appreciate the assistant from Mr. Shunfeng Li. Shenhe Fu and Yongyao Li appreciate the useful discussions from Profs. Jianying Zhou (Sun Yet-Sen University) and Boris A. Malomed (Tel-Aviv University). This work is supported by the National Natural Science Foundation of China
(Grant Nos.11575063, 61505265, 11374067), the Guangdong Provincial Science and Technology Plan (Grant No. 2014A050503064 and 2016A050502055), and the Special Funds for the Cultivation of Guangdong College Students Scientific and Technological Innovation (Grant No. pdjh2016b0077).

\bibliographystyle{plain}
\bibliography{apssamp}

\begin{thebibliography}{99}
\bibitem{Carson1964} L. J. Aplet and J. W. Carson, Appl. Opt. \textbf{3}, 544 (1964).
\bibitem{Kang2011} M. S. Kang, A. Butsch and P. St. J. Russell, Nat. Photonics \textbf{5}, 549 (2011).
\bibitem{Carlos1996} C. G. T. Palacios, G. I. Stegeman, and P. Baldi, Opt. Lett. \textbf{21}, 1442 (1996).
\bibitem{Gevorgyan2002} A. H. Gevorgyan, Tech. Phys. \textbf{47}, 1008 (2002).
\bibitem{Philip2007}R. Philip, M. Anija, C. S. Yelleswarapu, and D. V. G. L. N. Rao, Appl. Phys. Lett. {\bf 91}, 141118 (2007).
\bibitem{Fan2009} Z. Yu, S. Fan, Nat.Photonics \textbf{3}, 91  (2009).
\bibitem{Fan2012} L. Fan, J. Wang, L. T. Varghese, H. Shen, B. Niu, Y. Xuan, A, M. Weiner, M. Qi, Science \textbf{335}, 447 (2012).
\bibitem{Fan2015}Y. Shi, Z. Yu, and S. Fan, Nat. Photonics \textbf{9}, 388 (2015).
\bibitem{Wding2012}W. Ding, B. Luk'yanchuk, and C.-W. Qiu, Phys. Rev. A {\bf 85} 025806 (2012).
\bibitem{Xuyi2014}Y. Xu and A. E. Miroshnichenko, Phys. Rev. B {\bf 89}, 134306 (2014).
\bibitem{Lipson2009} S. Manipatruni, J. T. Robinson, and M. Lipson, Phys. Rev. Lett. \textbf{102}, 213903 (2009).
\bibitem{Feise2005}M. W. Feise, I. V. Shadrivov, and Y. S. Kivshar, Phys. Rev. E {\bf 71}, 037602 (2005).
\bibitem{Wang2013} D. Wang, H. Zhou, M. Guo, J. Zhang, J. Evers, and S. Zhu, Phys. Rev. Lett. \textbf{110}, 093901 (2013).
\bibitem{Zhang2014} Y. Zhang, D. Li, C. Zeng, Z. Huang, Y. Wang, Q. Huang, Y. Wu, J. Yu, and J. Xia, Opt. Lett. \textbf{39}, 1370 (2014).
\bibitem{Chang2014} L. Chang, X. Jiang, S. Hua, C. Yang, J. Wen, L. Jiang, G. Li, G. Wang, and M. Xiao, Nat. Photonics \textbf{8}, 524 (2014).
\bibitem{Khawaja2015} U. A. Khawaja, and A. A. Sukhorukov, Opt. Lett. {\bf 40}, 2719 (2015).
\bibitem{Hongji2015}H. Li, Z. Deng, J. Huang, S. Fu, and Y. Li, Opt. Lett. \textbf{40}, 2572 (2015).
\bibitem{BG1}J. T. Mok, C. M. de Sterke, I. C. M. Littler, and B. J. Eggleton, Nat. Phys. {\bf 2}, 775 (2006).
\bibitem{BG2}D. R. Neill, J. Atai, and B. A. Malomed, J. Opt. A {\bf 10}, 085105 (2008).
\bibitem{BG3}B. J. Eggleton, R. E. Slusher, C. M. de Sterke, P. A. Krug, and J. E. Sipe, Phys. Rev. Lett. {\bf 76}, 1627 (1996).
\bibitem{BG4}P. Y. P. Chen, B. A. Malomed, and P. L. Chu, Phys. Rev. E {\bf 71}, 066601 (2005).
\bibitem{BGT}D. N. Christodoulides and R. I. Joseph, Phys. Rev. Lett. {\bf 62}, 1746 (1989).
\bibitem{Atai2001}J. Atai, B. A. Malomed, Phys. Lett. A {\bf 284} 247 (2001).
\bibitem{Fu1}S. Fu, Y. Liu, Y. Li, L. Song, J. Li, B. A. Malomed, and J. Zhou, Opt. Lett. {\bf 38}, 5047 (2013).
\bibitem{Fu2}S. Fu, Y. Li, Y. Liu, J. Zhou, and B. A. Malomed, J. Opt. Soc. Am. B {\bf 32}, 534 (2015).
\bibitem{Caohui2012}H. Cao, J. Atai, X. Shu, and G. Chen, Opt. Exp. {\bf 20}, 12095 (2012).
\bibitem{SongLY}L. Song, S. Fu, Y. Liu, J. Zhou, V. G. Chigrinov, and I. C. Khoo, Opt. Lett. {\bf 38}, 5040 (2013).
\bibitem{Yikun2016}Y. Liu, Y. Wu, C. Chen, J. Zhou,1 T. Lin, and I. C. Khoo, Opt. Exp. {\bf 24}, 10458 (2016).
\bibitem{Hwang2004}J. Hwang, N. Y. Ha, H. J. Chang, B. Park, and J. W. Wu, Opt. Lett. {\bf 22}, 2644 (2004).
\bibitem{Hwang2005-2}J. Hwang and J. W. Wu, Opt. Lett. {\bf 30}, 875 (2005).
\bibitem{Hwang2005} J. Hwang, M. H. Song, B. Park, S. Nishimura, T. Toyooka, J. W. Wu, Y. Takanishi, K. Ishikawa and H. Takezoe, Nat. Material \textbf{4}, 383 (2005).
\bibitem{CBG}F. Ouellette, Opt. Lett. {\bf 12}, 847 (1987).
\bibitem{Ilchishin2011}I. P. Ilchishin, L. N. Lisetski, and T. V. Mykytiuk, Opt. Mater. Express {\bf 1}, 1484 (2011).
\end{thebibliography}

\end{document}